\documentclass[12pt]{article}
\usepackage{color}
\usepackage{amsmath}
\usepackage{cite}
\usepackage[pdftex]{graphicx}
\begin{document}    
\vspace*{1cm}

\renewcommand\thefootnote{\fnsymbol{footnote}}
\begin{center} 
  {\Large\bf Right-handed neutrino as a common progenitor of baryon number asymmetry 
and dark matter}
\vspace*{1cm}

{\Large Daijiro Suematsu}\footnote[1] {professor emeritus, ~e-mail:
suematsu@hep.s.kanazawa-u.ac.jp}
\vspace*{0.5cm}\\

{\it Institute for Theoretical Physics, Kanazawa University, 
Kanazawa 920-1192, Japan}
\end{center}
\vspace*{1.5cm} 

\noindent
{\Large\bf Abstract}\\
Cosmological and astrophysical observations suggest that both energy densities of baryon 
and dark matter take the same order values in the present Universe. We propose a scenario 
to give an answer for this problem in a scotogenic model and its extension. 
The model naturally provides dark matter candidates as its crucial ingredients to explain 
the small neutrino mass. If a neutral component of an inert doublet scalar plays a role of 
dark matter and leptogenesis occurs at TeV scales, both dark matter abundance and baryon 
number asymmetry could be explained with a same mother particle. Coincidence between  
the order of their energy densities might be understood through  such a background.

\newpage
\setcounter{footnote}{0}
\renewcommand\thefootnote{\alph{footnote}}

\section{Introduction}
Existence of dark matter (DM) suggested through various astrophysical and cosmological 
observations is one of unsolved problems in the standard model (SM) \cite{dmrev}. 
Since there is no corresponding particle in the SM, it can be a crucial clue 
to study new physics beyond the SM. 
A promising candidate is a weakly interacting massive particle called a WIMP.
If it is in the thermal equilibrium in the early Universe and is frozen out at a certain stage, 
it is known that the expected DM abundance could be naturally realized 
as its relic. 
However, such a particle has not yet been discovered in various kind of experiments, but
a lot of proposed models for it have 
been ruled out or constrained by them \cite{pdg,expb}.   

Another mysterious point on DM is cosmological coincidence of its abundance with
baryon number asymmetry, that is, why their energy densities take the same order 
values in the present Universe. 
The WIMP scenario cannot answer this problem since its abundance is usually considered 
to be explained through physics irrelevant to the baryon number asymmetry.
Asymmetric DM model \cite{asymdm} is a promising scenario which is motivated 
to give an answer for it. In this scenario, the DM abundance is explained as asymmetry of a 
certain conserved charge like the baryon number asymmetry.
If we consider an alternative possibility to explain it, a scenario may be constructed 
by assuming that both the baryon number asymmetry and the DM are caused from 
a same mother particle. 
In that case, the same physics could be relevant to them and then it could give an answer for
the problem. As such we study a class of model for the neutrino mass here. 
     
The scotogenic model \cite{scot} has been proposed to explain the small neutrino mass 
and the existence of DM, simultaneously. In this model,
the DM abundance is usually explained following the WIMP scenario. 
The model includes DM candidates as its important ingredients to generate 
the small neutrino mass. They are a lightest neutral component of an inert doublet scalar 
$\eta$ and a lightest right-handed neutrino.
If we choose the right-handed neutrino as the DM, a serious lepton-flavor violating problem 
appears \cite{lfv}. On the other hand, if we adopt a neutral component of
the $\eta$ as the DM, any serious phenomenological problem occurs, and then
it can be considered as a promising DM candidate in the model.  
Its relic abundance and various features have been extensively studied 
in a lot of papers \cite{scotdm,cross}. 

In this paper, we reconsider the DM abundance in this model from a view point of 
its relation to the baryon number asymmetry. In the scotogenic model, leptogenesis is known to 
generate baryon number asymmetry successfully \cite{basym,ks}.  
Since model parameters relevant to DM and leptogenesis could be almost independent 
in usually supposed cases, they have been studied separately. 
However, there could be an exceptional situation where they are closely related;
this situation seems not to have been noticed and has not been studied. 
In the present study, we focus our attention on a case where the model allows 
successful leptogenesis  at TeV scales as such a situation. 
There, we find that the usual estimation of the relic abundance of the DM should be modified. 
The DM abundance is considered to be explained as a cosmological relic which is 
neither a pure thermal relic nor an asymmetry of any conserved charge.   
It might give a new viewpoint for the cosmological coincidence between the DM 
and the baryon number asymmetry. 

Following parts are organized as follows. In section 2 we briefly review some features of 
the scotogenic model relevant to the present study. In section 3, after we overview 
leptogenesis and the DM physics in the model, we discuss a scenario in which the baryon number 
asymmetry and the DM abundance can be closely correlated. 
We show its realization in the model which can give the origin of the 
$CP$ phases in the PMNS matrix \cite{pmns}.  We summarize the paper in section 4. 
In the Appendix, we present a rough sketch of the derivation of the $CP$ phases 
in the PMNS matrix in the model.

\section{Scotogenic model}
The scotogenic model \cite{scot} is a simple extension of the SM with three right-handed neutrinos $N_j$ 
and an inert doublet scalar $\eta$. These new ingredients are assumed to have an
odd parity under imposed $Z_2$ symmetry although all the SM contents have
even parity. Relevant parts to these new ingredients in Lagrangian are given as 
\begin{eqnarray}
{\cal L}&\supset& \sum_{j=1}^3\left[\sum_{i=1}^3h_{i j}^\nu\bar\ell_{L_i}\eta N_j+M_{N_j}\bar N_jN_j^c 
 + {\rm h.c.}\right] +V,  \label{nulag} \\
V&=&m_\phi^2\phi^\dagger\phi+m_\eta^2\eta^\dagger\eta+
\lambda_1(\phi^\dagger\phi)^2
+\lambda_2(\eta^\dagger\eta)^2+\lambda_3(\phi^\dagger\phi)(\eta^\dagger\eta)
+\lambda_4(\phi^\dagger\eta)(\eta^\dagger\phi) \nonumber \\ 
&+&\left[\frac{\lambda_5}{2}(\eta^\dagger\phi)^2+{\rm h.c.}\right],
\label{npot}
\end{eqnarray}
where $\ell_{L_i}$ and $\phi$ stand for SM doublet leptons and a doublet Higgs scalar, 
respectively. The model can have a stable vacuum only if the potential $V$ in eq.~(\ref{npot})
satisfies the condition, 
\begin{equation}
\lambda_1,~ \lambda_2 >0, \quad \lambda_+> -2\sqrt{\lambda_1\lambda_2}, 
\label{stab}
\end{equation}  
where $\lambda_+$ is defined as $\lambda_+=\lambda_3+\lambda_4+\lambda_5$.

Since $\eta$ is assumed to have no vacuum expectation value (VEV), 
neutrino mass is forbidden at tree level but it can be generated by a VEV of the SM Higgs scalar 
through a one-loop diagram whose internal lines are composed of $N_j$ and $\eta$. 
Additionally, since $\eta$ gets no VEV, the imposed $Z_2$ symmetry remains 
as an exact symmetry and it guarantees the stability of the lightest $Z_2$ odd particle,
 which could be a cold DM candidate as long as it is neutral.
If $\lambda_4<0$ is satisfied, a lightest neutral component of $\eta$ can be such a candidate.
In the following study, $\lambda_5<0$ is assumed and then its real part $\eta_R^0$ 
is supposed to be DM.  Its mass $m_{\eta_R^0}$ has to satisfy  
$m_{\eta_R^0}< M_{N_1}$ where $M_{N_1}$ is mass of the lightest right-handed neutrino $N_1$.
In that case, $N_1$ can decay to the lepton through the Yukawa coupling
in eq.~(\ref{nulag}).  This decay could generate lepton number asymmetry,
which can be transformed to baryon number asymmetry through a sphaleron process \cite{spha,fy}.
It should be also noted that $\eta$ is also produced in this decay.
Thus, the lightest right-handed neutrinos could be a common mother particle of both 
the baryon number asymmetry and the DM in this model.    

The neutrino mass formula derived from the one-loop diagram is given as
\begin{eqnarray}
&&{\cal M}_{\nu_{ij}}=\sum_{k=1}^3 h_{ik}^\nu h_{jk}^\nu\Lambda_k, \nonumber \\
&&\Lambda_k=\frac{\lambda_5\langle\phi\rangle^2}{8 \pi^2M_{N_k}}
  \left[\frac{M_{N_k}^2}{M_\eta^2-M_{N_k}^2}
    \left(1+\frac{M_{N_k}^2}{M_\eta^2-M_{N_k}^2}
    \ln\frac{M_{N_k}^2}{M_\eta^2}\right) \right],
\label{numass}
\end{eqnarray}
where $M_\eta^2=m_\eta^2+(\lambda_3+\lambda_4)\langle\phi\rangle^2$.
This formula can explain the small neutrino mass required by neutrino 
oscillation data \cite{pdg} even for $N_j$ with the mass of TeV scales 
as long as $|\lambda_5|$ takes a sufficiently small value.\footnote{A lower 
bound on $|\lambda_5|$ can be derived through a possible inelastic scattering of $\eta_R^0$ 
with a nucleon in DM direct search experiments \cite{ks}.  } 
Moreover, if we note that only two right-handed neutrinos are necessary to derive 
two mass differences required to explain the neutrino oscillation data,
we find that the $N_1$ could be irrelevant to 
the small neutrino mass generation.
This suggests that the neutrino Yukawa coupling $h^\nu_{i1}$ can take a much smaller value
than others $h^\nu_{ij}~(j=2,3) $, which should be fixed so as to satisfy
the neutrino oscillation data through eq.~(\ref{numass}). 

Neutrino oscillation data suggest that the PMNS matrix which characterizes lepton 
flavor mixing could be described approximately through tribimaximal mixing \cite{otribi}. 
Although it cannot cause nonzero mixing angle $\theta_{13}$,
it can be modified by a mixing matrix for charged leptons even if the tribimaximal mixing 
matrix is assumed for the neutrino sector.\footnote{A concrete example of it is presented in \cite{cp} 
and also in the Appendix of this article. In the following study, parameters given there are used.} 
Tribimaximal mixing in the neutrino sector can be easily realized simply
by assuming the neutrino Yukawa couplings in eq.~(\ref{nulag}) to satisfy \cite{tribi},
\begin{equation}
h_{ek}=0, ~ h_{\mu k}=h_{\tau k}=h_k ~~(k=1,2); \quad h_{e3}=h_{\mu 3}=-h_{\tau 3}=h_3.
\label{tribi}
\end{equation}
Using this assumption, we can present examples which explain the neutrino oscillation 
data well. Here, we take $h_1$ to be sufficiently small like $O(10^{-8})$ so that 
$N_1$ is irrelevant to the neutrino mass determination as mentioned above.  
Taking account of this, if we apply to eq.~(\ref{numass}) the parameters\footnote{We note that 
$m_{\eta_R^0}\simeq M_\eta$ is satisfied for this small $|\lambda_5|$.},
\begin{eqnarray}
{\rm (a)} ~~\lambda_5=-10^{-5}, \quad m_{\eta_R^0}=2\times 10^3~{\rm GeV},
\quad M_{N_2}=6\times 10^4~{\rm GeV}, \quad  M_{N_3}=7\times 10^4~{\rm GeV}, \nonumber \\
{\rm (b)} ~~\lambda_5=-10^{-5}, \quad m_{\eta_R^0}=2\times 10^3~{\rm GeV},
\quad M_{N_2}=6\times 10^3~{\rm GeV}, \quad  M_{N_3}=7\times 10^3~{\rm GeV}, 
\label{scotpara}
\end{eqnarray}
the mass differences required to explain the neutrino oscillation data are found to be 
realized for the couplings  
\begin{eqnarray}
{\rm (a)}\quad h_2=1.16\times 10^{-2}, \qquad h_3=4.15\times 10^{-3}, \nonumber \\
{\rm (b)}\quad h_2=6.87\times 10^{-3}, \qquad h_3=2.37\times 10^{-3}.
\label{h23}
\end{eqnarray}
We use them in the following study. 

\section{Leptogenesis and dark matter abundance}
\subsection{Low-scale leptogenesis}
This model makes leptogenesis \cite{fy} work well to generate the baryon number 
asymmetry in the same way as the type-I seesaw model \cite{seesaw}. 
Sufficient lepton number asymmetry could be produced as a seed of the baryon 
number asymmetry through the out-of-equilibrium decay of $N_1$ \cite{ks}.
Moreover, it could happen at much lower scales than the type-I seesaw model if $N_1$ 
is in the thermal equilibrium \cite{lowlept}.
However, it is difficult to make it in the thermal equilibrium only by the neutrino Yukawa 
coupling $h^\nu_{i1}$ in eq.~(\ref{nulag}) in a consistent way with successful leptogenesis 
as suggested in \cite{ks}. To overcome this difficulty,
interactions, which can produce a sufficient amount of $N_1$ to reach its equilibrium 
value, are proposed in some extended frameworks of the scotogenic model in connection with 
various problems of the SM, for example, inflation \cite{infnonth, infth}, 
right-handed neutrino mass \cite{rightn1,rightn2}, and $CP$ violation \cite{cp0,cp}.  
For a while, we assume that $N_1$ is in the thermal equilibrium through certain interactions.
Study of the coincidence problem based on a concrete interaction is given later.  

The $CP$ asymmetry $\varepsilon$ in the decay $N_1\rightarrow \ell_{L_i}\eta^\dagger$ 
is expressed as \cite{cpasym}
\begin{eqnarray}
\varepsilon&\equiv&\frac{\sum_i[\Gamma(N_1\rightarrow \ell_{L_i}\eta^\dagger)-
\Gamma(N_1^c\rightarrow \bar\ell_{L_i}\eta)]}{\sum_i\Gamma(N_1\rightarrow \ell_{L_i}\eta^\dagger)}
\nonumber \\
&=&\frac{1}{8\pi}\sum_{j=2,3}
{\rm Im}\left[\frac{(\sum_i h_{i1}^\nu h_{ij}^\nu)^2}
{\sum_i h_{i1}^{\nu 2}}\right]F\left(\frac{M_{N_j}^2}{M_{N_1}^2}\right),
\label{asymp}
\end{eqnarray}
where $F(x)=\sqrt{x}[1-(1+x)\ln\frac{1+x}{x}]$.
Since the dependence on $h^\nu_{i1}$ in this asymmetry $\varepsilon$ can be  
suppressed under a suitable flavor structure,  $h^\nu_{i1}$ can be assigned a much smaller value 
compared with $h^\nu_{i2}$ and $h^\nu_{i3}$ keeping a value of $\varepsilon$ to be a 
magnitude required for successful leptogenesis.
On the other hand, a small value of $h^\nu_{i1}$ makes 
the decay of $N_1$ delay so that the washout of generated lepton 
number asymmetry could be ineffective when its substantial decay starts. 
These could make successful leptogenesis possible for the $N_1$ with the TeV scale mass 
in the scotogenic model \cite{lowlept,rightn2,infth,cp0,cp}.

The decay of $N_1$ is expected to start around the temperature $T_L$, which satisfies 
a condition $\Gamma_{N_1}^D=H(T_L)$, where $\Gamma_{N_1}^D$ and $H(T)$ are 
the decay width of $N_1$ and the Hubble parameter at the temperature $T$, respectively. 
They are expressed as
\begin{equation}
\Gamma_{N_1}^D=\sum_{i=1}^3\frac{h_{i1}^{\nu 2}}{8\pi}M_{N_1}\sqrt{1-\frac{M_\eta^2}{M_{N_1}^2}}, 
\qquad H(T)=\left(\frac{\pi^2}{90}g_\ast\right)^{1/2}\frac{T^2}{M_{\rm pl}},
\label{hub}
\end{equation}
where $g_\ast$ is relativistic degrees of freedom at $T$ and $M_{\rm pl}$ is a reduced Planck mass.
If we use the tribimaximal assumption given in eq.~(\ref{tribi}) for the neutrino Yukawa couplings, 
the temperature $ T_L$ can be estimated as
\begin{equation}
\frac{T_L}{M_{N_1}}=
5.1\times \left(\frac{h_1}{10^{-8}}\right)\left(\frac{10^4~{\rm GeV}}{M_{N_1}}\right)^{1/2},
\label{decay}
\end{equation}
where $g_\ast=116$ is used.

\begin{figure}[t]
\begin{center}
\includegraphics[width=7cm]{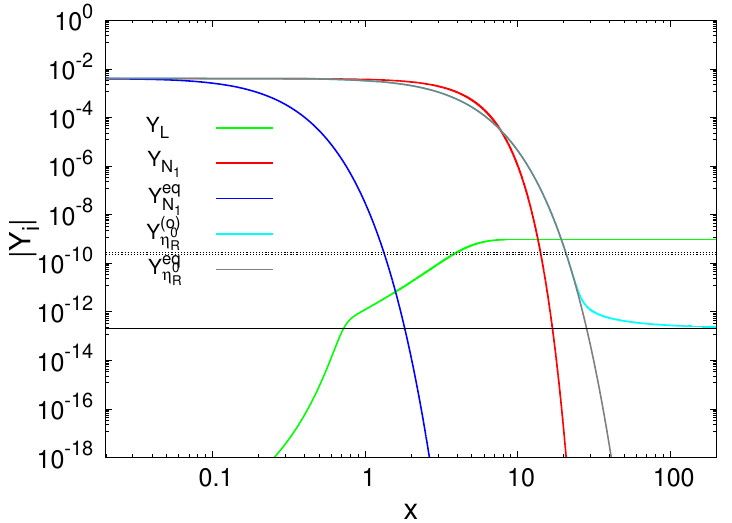}
\hspace*{5mm}
\includegraphics[width=7cm]{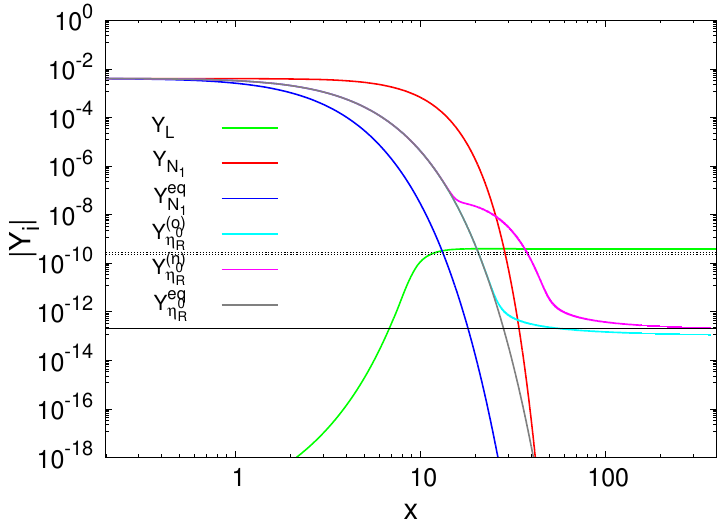}
\end{center}

\footnotesize{{\bf Fig.~1}~~ Evolution of the lepton number asymmetry $Y_L$, the lightest right-handed 
neutrino number density $Y_{N_1}$ and the $\eta_R^0$ number density $Y_{\eta_R^0}$ 
is displayed as a function of a dimensionless parameter $x(\equiv m_{\eta_R^0}/T)$. 
$N_1$ contribution is not taken into account in $Y_{\eta_R^0}^{(o)}$ but 
it is taken into account in $Y_{\eta_R^0}^{(n)}$. $Y_i$ with superscript ``eq'' stands for 
its equilibrium value. In the left panel for the case (a), 
$h_1=2.0\times 10^{-8}$ and $M_{N_1}=3.1\times 10^4~{\rm GeV}$ are used.
In the right panel for the case (b), $h_1=5.4\times 10^{-8}$ and $M_{N_1}=3.1\times 10^3~{\rm GeV}$ 
are used. Horizontal lines represent values of $Y_L$ (dotted) 
and $Y_{\eta_R^0}$ (solid) which are required to explain the abundance of the baryon 
number asymmetry and the DM abundance in the present universe, respectively.}
\end{figure}

Here, we define $Y_i$ as $Y_i\equiv \frac{n_i}{s}$ by using the number 
density $n_i$ of a particle species $i$ and the entropy density $s$.
Lepton number asymmetry is expressed by $Y_L\equiv \frac{n_\ell-n_{\bar\ell}}{s}$,
where $n_\ell$ and $n_{\bar \ell}$ are the number density of leptons and antileptons, 
respectively.
If washout processes of the lepton number asymmetry $Y_L$ decouple at a temperature 
$T_F$ and $T_L<T_F$ is satisfied,
the $Y_L$ generated through the $N_1$ decay 
is not affected by the washout effect which is caused by the inverse decay of
the right-handed neutrinos and 2-2 scatterings mediated by $N_{2,3}$.\footnote{If the lepton 
asymmetry is generated at the temperature lower than the $\eta$ mass as a result of a small $h_1$, 
it can escape the washout by these processes since they 
could be sufficiently suppressed by the Boltzmann factor.
It is confirmed through the numerical study shown in Fig.~3 which shows that $T_L$ and $T_F$ satisfy 
$T_L<T_F~{^<_\sim}~m_{\eta_R^0}/10$ as expected.} 
In such a case, the lepton number asymmetry at $T_L$ can be roughly 
estimated by using $\varepsilon$ in eq.~(\ref{asymp}) as 
$Y_L=\varepsilon Y_{N_1}^{\rm eq }(T_L)$ where $Y^{\rm eq}_{N_1}(T_L)$ takes 
the equilibrium number density of the relativistic particle as 
$Y^{\rm eq}_{N_1}(T_L)=O(10^{-3})$.
Since $T_L< M_{N_1}$ has to be satisfied at least, eq.~(\ref{decay}) shows 
\begin{equation}
h_1<6.1\times 10^{-7}\left(\frac{M_{N_1}}{10^4~{\rm GeV}}\right)^{1/2}.
\label{h1cond1}
\end{equation}
The required baryon number asymmetry can be generated for $Y_L=O(10^{-10})$ and then  
$\varepsilon$ should have a value of $O(10^{-7})$. 
Eq.~(\ref{asymp}) shows that such an $\varepsilon$ is consistent with the values 
of $h_{2,3}$ given in eq.~(\ref{h23}) as long as a maximal $CP$ violation is assumed. 

The generated lepton number asymmetry is converted to the baryon number asymmetry 
through the sphaleron process which is considered to be in the thermal equilibrium 
at temperature higher than 100 GeV \cite{spha}.  Since the lepton number violating $N_1$ decay
has to occur at a higher temperature than it, this imposes that $\Gamma_{N_1}^D>H(T)$ 
should be satisfied at $T=100$~ GeV. This condition can be expressed as
\begin{equation}
h_1>6.1\times 10^{-9}\left(\frac{10^4~{\rm GeV}}{M_{N_1}}\right)^{1/2}.
\label{h1cond2}
\end{equation}
If we impose that leptogenesis occurs successfully at TeV regions, we find 
from eqs.~(\ref{h1cond1}) and (\ref{h1cond2}) that $h_1$ should take a value 
of $O(10^{-8})$ for $M_{N_1}$, which is larger than $m_{\eta_R^0}$ assumed in eq.~(\ref{scotpara}). 

To examine these qualitative arguments, we solve the Boltzmann equations 
for $Y_L$ and $Y_{N_1}$ given in \cite{ks} numerically 
by assuming that $N_1$ is initially in the thermal equilibrium and using the the 
parameters given in eqs.~(\ref{scotpara}) and (\ref{h23}).
The results for both cases (a) and (b)  are given in the left and right panels of Fig.~1, 
respectively. Both panels show that the $N_1$ decay delays and $Y_{N_1}$ 
keeps its relativistic value until the temperature $T$ reaches the value which satisfies
$T\ll M_{N_1}$.
Since $Y_L$ is found to be realized as $\varepsilon Y_{N_1}^{\rm eq}(T_L)$ from it,  
the above discussion can be justified quantitatively.  
The sufficient lepton number is found to be produced in both cases.

\subsection{DM abundance}
We consider the $\eta_R^0$ abundance as DM in this model.
The relic abundance of $\eta_R^0$ is determined by solving Boltzmann equation for it.
If we define a dimensionless parameter $x$ as 
$x\equiv m_{\eta_R^0}/T$, Boltzmann equations for $Y_{\eta_R^0}$ and $Y_{N_1}$ 
can be written as 
\begin{eqnarray}
&&\frac{dY_{\eta_R^0}}{dx}=-\frac{s(m_{\eta_R^0})}{H(m_{\eta_R^0})x^2}
\langle\sigma v\rangle(Y_{\eta_R^0}^2-Y_{\eta_R^0}^{{\rm eq}^2})
+\frac{2x}{H(m_{\eta_R^0})}\Gamma^D_{N_1}(Y_{N_1}-Y_{N_1}^{\rm eq}), 
\label{bolt1} \\
&&\frac{dY_{N_1}}{dx}=-\frac{x}{H(m_{\eta_R^0})}
\Gamma^D_{N_1}(Y_{N_1}-Y_{N_1}^{\rm eq}),
\label{bolt2}
\end{eqnarray}
where $H(m_{\eta_R^0})$ and $s(m_{\eta_R^0})$ represent the Hubble parameter and the
entropy density at $T=m_{\eta_R^0}$, respectively.
$\langle\sigma v\rangle$ is a thermally averaged annihilation cross section of $\eta_R^0$ 
and $\Gamma_{N_1}^D$ is the decay width of $N_1$ given in eq.~(\ref{hub}).
We take account of an effect of the $N_1$ decay to $\eta$ as the second term 
in the right-hand side of eq.~(\ref{bolt1}). 
Here, we define $Y_{\eta_R^0}^{(n)}$ and  $Y_{\eta_R^0}^{(o)}$ as the solution of 
eq.~(\ref{bolt1}) with the second term and the one without it, respectively.
Since this term can be safely neglected for the case $M_{N_1} \gg m_{\eta_R^0}$, 
we can expect $Y_{\eta_R^0}^{(n)}=Y_{\eta_R^0}^{(o)}$ there and DM is considered as 
thermal relic. We consider such a case first. 

In that case, the present abundance of $\eta_R^0$ is fixed through its equilibrium density 
at temperature $T_D$ where its annihilation processes are frozen out \cite{dmrelic,coan}.  
We can estimate $T_D$ by using its thermally averaged annihilation cross section 
$\langle\sigma v\rangle$ and Hubble parameter $H(T)$ through a condition
$2n_{\eta_R^0}^{\rm eq}(T_D)\langle\sigma v\rangle =H(T_D)x_D$ 
as found from eq.~(\ref{bolt1}). 
 Using $H(T)$ given in eq.~(\ref{hub}), $x_D$ can be determined by solving 
the above mentioned condition as
\begin{equation}
x_D=\ln\frac{0.384 g M_{\rm pl} \langle\sigma v\rangle m_{\eta_R^0}}
{(g_\ast x_D)^{1/2}},
\label{dtemp}
\end{equation}
where $g$ is an internal degree of freedom of $\eta_R^0$. 
Since $Y_{\eta_R^0}$ converges to a constant value $Y_{\eta_R^0}^\infty$ at $x>x_D$, 
the present DM abundance can be 
expressed as $\Omega_{\eta_R^0}=m_{\eta_R^0}Y_{\eta_R^0}^\infty s_0/\rho_0$ where
$\rho_0$ and $s_0$ are the energy density and the entropy density 
in the present Universe.  They are given as $\rho_0=3M_{\rm pl}^2 H_0^2$ and 
$s_0=2.27\times 10^{-38}$~GeV$^3$.
If we solve eq.~(\ref{bolt1}) by taking account of $Y_{\eta_R^0}>Y_{\eta_R^0}^{\rm eq}$ 
at $x\gg x_D$ and also $Y_{\eta_R^0}^\infty<Y_{\eta_R^0}(x_D)$, 
we find that $\Omega_{\eta_R^0}$ can be approximately expressed as
\begin{equation}
\Omega_{\eta_R^0} h^2=\frac{Y^\infty_{\eta_R^0} m_{\eta_R^0}s_0}{3M_{\rm pl}^2H_0^2/h^2}
=\frac{2.13\times 10^8~{\rm GeV}}
{\sqrt{g_\ast}M_{\rm pl}\int_{x_D}^\infty\frac{\langle\sigma v\rangle}{x^2}dx},
\label{dabund}
\end{equation} 
where we use $H_0=2.13\times 10^{-42}h$ GeV.
Applying it to the present observational result $\Omega_{DM}h^2=0.12$,
we find that $Y_{\eta_R^0}^\infty=2.13 \times 10^{-13}\left(2000~{\rm GeV}/m_{\eta_R^0}\right)$ 
should be satisfied. 
It constrains the parameters $\lambda_+$ and $\lambda_3$ in the potential (\ref{npot})
which determine the annihilation cross section $\langle\sigma v\rangle$.
Since the masses of the components of $\eta$ are almost degenerate, 
coannihilation \cite{coan} among them should be taken into account to estimate 
$\langle\sigma v\rangle$ \cite{ks,cross}.

\begin{figure}[t]
\begin{center}
\includegraphics[width=7cm]{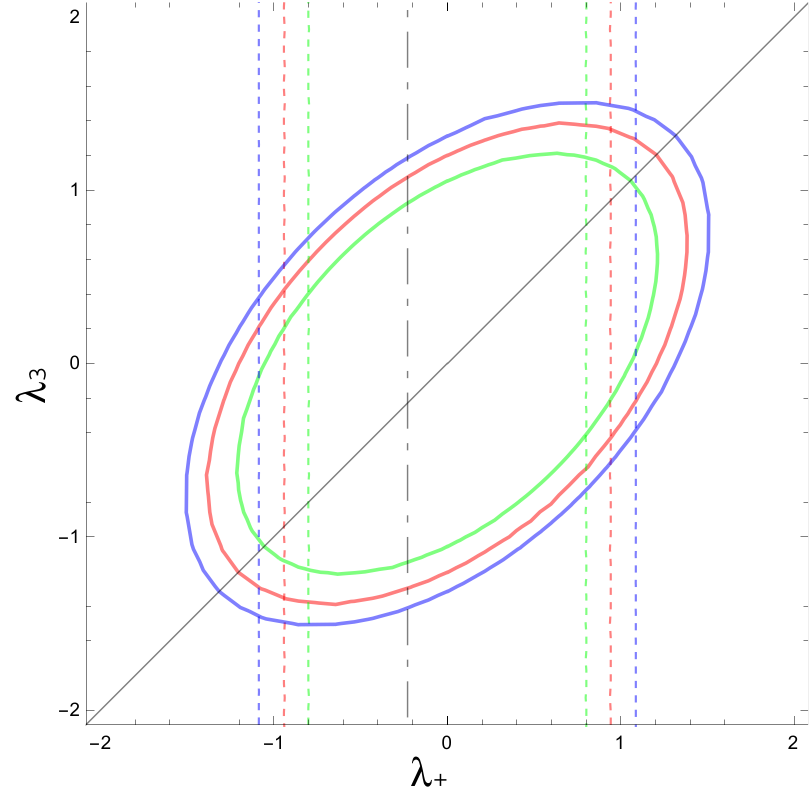}
\end{center}

\footnotesize{{\bf Fig.~2}~~ Contours of the $\eta_R^0$ relic density 
$\Omega_{\eta_R^0}h^2=0.12$  are plotted by colored solid lines  in
the $(\lambda_+, \lambda_3)$ plane. Each contour corresponds to the one with 
$m_{\eta_R^0}$=2200~GeV (blue), 2000~GeV (red), and 1800~GeV (green), respectively.
An upper bound on $|\lambda_+|$ based on the direct search of XENONnT is represented
by vertical dashed lines with the same color as the one of the relic density 
for each mass. A diagonal black solid line and a vertical black dash-dotted line represent 
the condition $\lambda_4=0$ and $\lambda_+=-2\sqrt{\lambda_1\lambda_2}$ 
with $\lambda_2=0.1$, respectively. The allowed region is restricted into an 
upper-triangle region surrounded by these lines.}
\end{figure}

In Fig.~2, we draw contours of $\Omega_{\eta_R^0}h^2=0.12$ for typical values of $m_{\eta_R^0}$ 
in the $(\lambda_+, \lambda_3)$ plane.
In this plane, we have to take account of both the stability condition given in eq.~(\ref{stab}) and 
$\lambda_4=\lambda_+-\lambda_3-\lambda_5<0$ which is a necessary condition for 
$\eta_R^0$ to be lighter than the charged components.
Combining them, we find that only points on the contours
included in a triangle region of the upper-right quadrant in the $(\lambda_+,\lambda_3)$ plane 
are allowed. 

On the other hand, the same parameters are also constrained by the present results of
DM direct search experiments. In this model, $\eta_R^0$-nucleon 
elastic scattering is caused by the $t$-channel Higgs exchange. 
Its cross section can be expressed as
\begin{equation}
\sigma_N=\frac{\lambda_+^2}{8\pi}\frac{\bar f_N^2}{m_{\eta_R^0}^2}\frac{m_N^4}{m_h^4},
\label{direct}
\end{equation}
where $\bar f_N$ is a coupling between the Higgs scalar and a nucleon.
Masses of the nucleon and the Higgs boson are represented by  $m_N$ and $m_h$, respectively.
A present most stringent bound on $\sigma_N$ through the direct search experiment 
is given by the XENONnT experiment \cite{xenon}.
Since it gives an upper bound on $|\lambda_+|$, as found from eq.~(\ref{direct}),
a region included in a band sandwiched with dotted vertical lines fixed for each 
$m_{\eta_R^0}$ is allowed.
These suggest that only restricted points in the $(\lambda_+,\lambda_3)$ plane
can be consistent with the present data for the DM abundance.
Future direct detection experiments and collider experiments might find a $\eta_R^0$ signature 
in this parameter region.

\subsection{Coincidence of  DM and baryon number asymmetry}
We reconsider realization of the DM abundance in relation to the baryon number
asymmetry generated through leptogenesis. 
Since the required relic abundance is known to be realized for $x_D\sim 25$ 
in the WIMP scenario, $T_D<100~{\rm GeV} <T_L$ can be satisfied for $\eta_R^0$ 
whose mass is less than 3 TeV.
If $M_{N_1} \gg m_{\eta_R^0}$ is satisfied, $Y_{\eta_R^0}^{\rm eq}(x_D)\gg Y_{N_1}^{\rm eq}(x_D)$
 is expected and then $Y_{\eta_R^0}^{(n)}=Y_{\eta_R^0}^{(o)}$ is guaranteed.
So that the estimation of the relic abundance of $\eta_R^0$ in the previous section can be 
justified. 

The solution of the Boltzmann equation for $Y_{\eta_R^0}$ 
added  in the left panel of  Fig.~1 proves it.  
In this calculation, $\lambda_+$ and $\lambda_3$ are fixed to 0.66 and 1.56, 
which are the consistent values with the contour for $m_{\eta_R^0}=2$ TeV shown in Fig.~2.
There is no contribution to $Y_{\eta_R^0}$ from the $N_1$ decay 
and then $Y_{\eta_R^0}^{(o)}$ is found to be equal to $Y_{\eta_R^0}^{(n)}$.
The annihilation processes of $\eta_R^0$ are frozen out at $x_D\sim 26$ where 
$Y_{N_1}$ becomes much smaller than $Y^{\rm eq}_{\eta_R^0}$.
After that, $Y_{\eta_R^0}$ converges to a constant value which gives $\Omega_{\eta_R^0}h^2=0.12$. 
It suggests that $Y_{\eta_R^0}$ is determined irrelevantly from 
the determination of $Y_L$. It is generally expected for the case where $N_1$ is much heavier 
than $\eta_R^0$. 

Next, we consider a case where $Y^\infty_{\eta_R^0}$ is determined through 
a different way from the above case.
It can happen in a situation where $h_1$ takes a small value and also $N_1$ has 
the similar mass to $\eta_R^0$.
The condition for $h_1$ required by the leptogenesis is given by 
eqs.~(\ref{h1cond1}) and (\ref{h1cond2}).  
If the Yukawa coupling $h_1$ takes a value such as $H(T_D)~{^>_\sim}~ \Gamma_{N_1}^D$,
the $N_1$ abundance could be larger than $Y_{\eta_R^0}^{\rm eq}(x_D)$ which 
saturates the required DM abundance.
To escape such a disastrous situation,  $h_1$ has to satisfy 
\begin{equation} 
h_1>4.9\times 10^{-9}\left(\frac{10^4~{\rm GeV}}{M_{N_1}}\right)^{1/2}.
\label{h1cond3}
\end{equation}
This is consistent with eq.~(\ref{h1cond2}). If these conditions are satisfied and
$Y_{\eta_R^0}^{\rm eq}(x_D)$ is smaller than the value required by the DM abundance,
 the decay of $N_1$ could give a crucial contribution to the relic density of $\eta_R^0$
 since $Y_{N_1}(x_D)~{^>_\sim}~ Y_{\eta_R^0}^{\rm eq}(x_D)$ could occur.

We can estimate $Y_{N_1}(x_D)$ by solving eq.~(\ref{bolt2}) as
\begin{equation}
Y_{N_1}(x_D)=Y_{N_1}(x_i) \exp\left[-\frac{\Gamma_{N_1}^D}{2H(m_{\eta_R^0})}(x_D^2-x_i^2)\right].
\label{yn1}
\end{equation} 
Since $Y_{N_1}(x_i)$ is of $O(10^{-3})$ at $T_i=M_{N_1}$ which stands for 
$x_i=m_{\eta_r^0}/M_{N_1}(<1)$,
the exponential factor in eq.~(\ref{yn1}) can be rewritten as
\begin{equation}
\exp\left[-\frac{\Gamma_{N_1}^D}{2H(m_{\eta_R^0})}x_D^2\right]=
\exp\left[-4.3\times 10^{12}\left(\frac{x_D}{25}\right)^2
h_1^2M_{N_1}\sqrt{1-\left(\frac{m_{\eta_R^0}}{M_{N_1}}\right)^2}\right],
\label{exp}
\end{equation}
where we use $m_{\eta_R^0}=2000$~GeV. The required DM abundance can be 
realized for $Y_{\eta_R^0}(x_D)=O(10^{-12})$ as found from the left panel of Fig.~1. 
If we impose $Y_{N_1}(x_D)=O(10^{-12})$, eqs. (\ref{yn1}) and (\ref{exp}) suggest that
the $\eta_R^0$ produced by the $N_1$ decay at $x>x_D$ could supply a substantial 
part of the DM abundance for
\begin{equation}
h_1^2M_{N_1}\sqrt{1-\left(\frac{2000}{M_{N_1}}\right)^2}= O( 10^{-12}).
\label{cond}
\end{equation}
This relation is consistent with conditions (\ref{h1cond1}) and (\ref{h1cond2}).

If $h_1$ and $M_{N_1}$ take values following the condition (\ref{cond}), 
$Y_{N_1}(x_D)~{^>_\sim}~Y_{\eta_R^0}(x_D)$ can be realized 
although $N_1$ is heavier than $\eta_R^0$.
The $N_1$ decay which causes the lepton number asymmetry could also generate 
a dominant part of the relic $\eta_R^0$ for parameters $\lambda_+$ and $\lambda_3$ 
which are ruled out in the ordinary estimation of the relic $\eta_R^0$.
The present abundance of $\eta_R^0$ should be estimated based on
\begin{equation}
Y_{\eta_R^0}(x_D)= Y_{\eta_R^0}^{\rm eq}(x_D)+2Y_{N_1}(x_D),
\end{equation}
since all components of $\eta$ produced through the $N_1$ decay
finally come to $\eta_R^0$. 
As the Boltzmann equation for $Y_{\eta_R^0}$,
we have to use eq.~(\ref{bolt1}) which takes account of the $N_1$ decay to $\eta$. 

To examine the above observation quantitatively, we solve Boltzmann equations 
(\ref{bolt1}) and (\ref{bolt2}) numerically assuming that $N_1$ is initially in the thermal 
equilibrium. 
Although the mass difference between $\eta_R^0$ and $N_1$ is small,  
their coannihilation can be neglected in $\langle\sigma v\rangle$ because their coupling 
$h_1$ is small enough.
In the right panel of Fig.~1, the evolution of both $Y_{\eta_R^0}^{(n)}$ and $Y_{\eta_R^0}^{(o)}$ 
obtained as the solutions of the Boltzmann equations in the case (b) is plotted. 
In this calculation, $\lambda_+$ and $\lambda_3$ are fixed to 0.66 and 2.14, respectively.
These make $Y_{\eta_R^0}^{(o)}$ coming from the thermal $\eta_R^0$ smaller than 
the required value $\Omega_{\eta_R^0}h^2=0.12$. It occupies about 50 \% of the total
and a remaining part is supplied through the $N_1$ decay.
The figure shows that both the sufficient baryon number asymmetry and the DM abundance 
are simultaneously realized through the $N_1$ decay.

Comparison of both panels in Fig.~1 clarifies features of the present case.
 In the left panel which corresponds to the usually supposed case, 
the $\eta_R^0$ abundance is realized by the freeze-out of the thermal $\eta_R^0$ 
and the yields from the $N_1$ decay is irrelevant.
The baryon number asymmetry and the DM abundance is explained based 
on irrelevant physics.
On the other hand, we can find that the $N_1$ decay plays a crucial role to determine 
the $\eta_R^0$ abundance in the right panel.  
In this example, the contribution from the thermal $\eta_R^0$ is only $50 \%$ and 
the remaining one is caused by the $N_1$ decay.  
It suggests that the $\eta_R^0$ yielded through the $N_1$ decay could supply the substantial part 
of the $\eta_R^0$ abundance. 
The coincidence of the baryon number density and the dark matter density 
in the present Universe could be explained naturally there 
since common parameters relevant to the $N_1$ decay control them simultaneously.
 
\subsection{A feasible model for the scenario}
Finally, in order to show that this scenario works in a realistic way,   
we adopt a well motivated model with possible interactions which can generate $N_1$ 
in the thermal equilibrium even if the Yukawa coupling $h_1$ is too small
to produce it effectively.  
The model has been proposed to give a prospect to the $CP$ issues in the SM \cite{cp}.
It can solve the strong $CP$ problem \cite{strongcp} through the Nelson-Barr mechanism 
\cite{nb,bbp} and 
give the origin of $CP$ phases in the PMNS and CKM matrices \cite{km,pmns}. 

This model is an extension of the scotogenic model with
a singlet scalar $S$ and vectorlike fermions $D_{L,R}$ and $E_{L,R}$, where
$D_{L,R}$ and $E_{L,R}$ are down-type singlet quarks and charged leptons, respectively.
Their Yukawa couplings are given as
\begin{eqnarray}
&& \sum_{k=1}^3\left[ 
(y_k^dS+\tilde y_k^dS^\dagger)\bar D_L d_{R_k}  + (y_k^eS+\tilde y_k^eS^\dagger)\bar E_L e_{R_k} 
+ (y_{N_k}S+\tilde y_{N_k}S^\dagger)\bar N_kN_k^c\right]  \nonumber \\
&&+(y_ES +\tilde y_E S^\dagger)\bar E_LE_R +{\rm h.c.}, 
  \label{qlag} 
\end{eqnarray}
where $d_{R_k}$ and $e_{R_k}$ are the singlet down-type quark and the singlet charged
lepton in the SM, respectively. Since $CP$ invariance is assumed in the model, 
all coupling constants in eq.~(\ref{qlag}) are considered to be real.
If the singlet scalar $S$ gets a VEV as $\langle S\rangle=\frac{1}{\sqrt 2}ue^{i\rho_0}$,
spontaneous $CP$ violation is caused. 
Complex phases due to this $CP$ violation could be brought about 
in the PMNS and CKM matrices through the mixing between the SM fermions and 
vectorlike fermions which is caused by these interactions.\footnote{A rough sketch of 
this scenario is given in the Appendix.}  
The last term in the first line of (\ref{qlag}) induces the mass of the right-handed neutrino $N_k$. 
If we redefine $N_k$ to make its mass real, $M_{N_k}$ and $\Lambda_k$
in eq.~(\ref{numass}) can be expressed as   
\begin{eqnarray}
&&M_{N_k}=(y_{N_k}^2+\tilde y_{N_k}^2+2y_{N_k}\tilde y_{N_k}\cos 2\rho_0)^{1/2}u, \nonumber \\
&&\Lambda_k=|\lambda_k|e^{i\theta_k}, \qquad 
 \tan\theta_k=\frac{y_{N_k}-\tilde y_{N_k}}{y_{N_k}+\tilde y_{N_k}}\tan\rho_0.
\label{theta}
\end{eqnarray}
This $\theta_k$ determines the $CP$ violation in the $N_1$ decay 
and then it fixes the $CP$ asymmetry $\varepsilon$ given in eq.~(\ref{asymp}). 

\begin{figure}[t]
\begin{center}
\includegraphics[width=7cm]{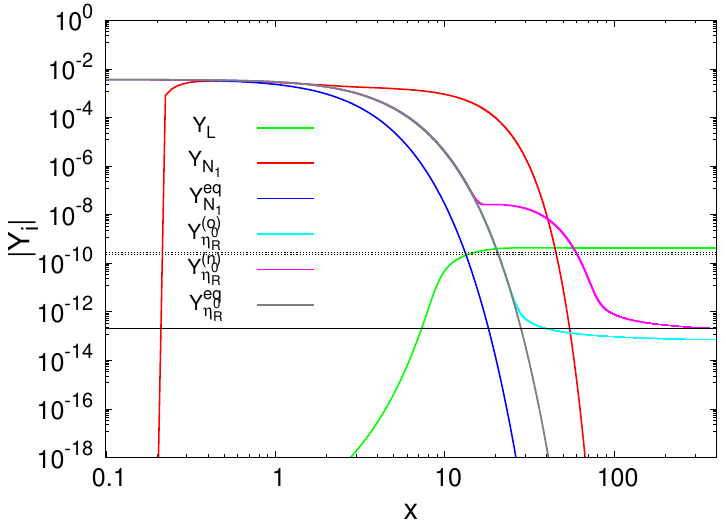}
\hspace*{5mm}
\includegraphics[width=7cm]{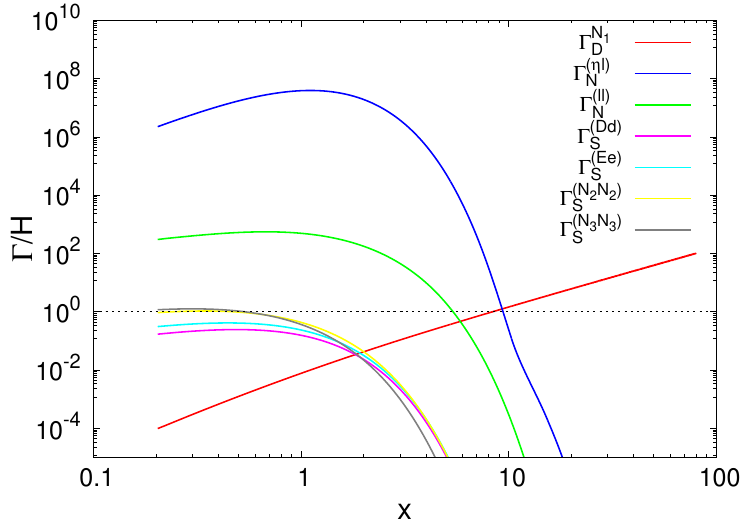}
\end{center}

\footnotesize{{\bf Fig.~3}~~Left: Evolution of the lepton number asymmetry $Y_L$ and 
the $\eta_R^0$ number density $Y_{\eta_R^0}$.  In the calculation, $h_1=3.4\times 10^{-8}$
and $M_1=3.1\times 10^3$ GeV for the case (b) are used. Horizontal lines represent 
the required values of $Y_L$ (dotted) and $Y_{\eta_R^0}$ (solid) to explain the  
baryon number asymmetry and the DM abundance, respectively.
Right: Evolution of the relevant reaction rates included in the Boltzmann equations.
$\Gamma_f^{(ab)}$ stands for the reaction rate of the scattering $ab\rightarrow N_1N_1$
mediated by $f$. The washout of the lepton number asymmetry $Y_L$ is caused by
$\Gamma^{(\eta\ell)}_N$ and $\Gamma^{(\ell\ell)}_N$.}
\end{figure}

In the context of this paper, we should especially note that these interactions can cause 
scatterings $\bar D_Ld_{R_k}\rightarrow \bar N_1N_1^c$, 
$\bar E_Le_{R_k}\rightarrow \bar N_1N_1^c$, 
$\bar E_LE_R\rightarrow \bar N_1N_1^c$, and
$\bar N_kN_k^c\rightarrow \bar N_1N_1^c$ through the exchange of $S$.   
If the mass of the vectorlike fermions is smaller than the reheating temperature,
they could be in the thermal equilibrium through the SM gauge interactions.
Heavier right-handed neutrinos $N_{2,3}$ could also be in the thermal equilibrium effectively 
through their neutrino Yukawa interactions at the temperature less than $O(10^8)$ GeV 
if their couplings take the values given in eq.~(\ref{h23}) \cite{rightn1,rightn2}.
In such a case, these processes can produce $N_1$ effectively to reach its 
equilibrium density.
Thus, if its mass $M_{N_1}$ takes a similar value to $m_{\eta_R^0}$ as in the case (b) 
shown in the right panel of Fig.~1, we can expect that the $N_1$ decay generates the lepton 
number asymmetry sufficiently and it also contributes to the relic abundance 
of $\eta_R^0$ substantially.\footnote{We should note that a scalar interaction 
$\kappa_{S\eta}S^\dagger S\eta^\dagger\eta$ is not forbidden in the model, which could give an 
additional source of $\eta_R^0$ through the scattering. 
However, since this process is effective at the temperature higher than
$m_{\eta_R^0}$, it does not affect the present calculation of the DM abundance.
It should be also noted that any change in the neutrino mass formula (\ref{numass}) is not caused 
by this interaction since the one-loop neutrino mass depends on the squared mass difference 
of $\eta_R^0$ and $\eta_I^0$ which is not changed by this interaction.}

For its quantitative check, we solve the Boltzmann equations for $Y_{N_1}, Y_{\eta_R^0}$, 
and $Y_L$ by taking account of these scattering processes. 
Eq.~(\ref{bolt2}) should be modified by introducing the right-hand side additional terms
\begin{equation}
-\frac{s(m_{\eta_R^0})}{H(m_{\eta_R^0})x^2}\sum_\alpha
\langle\sigma v\rangle_\alpha(Y_{N_1}^2-Y_{N_1}^{{\rm eq} 2}),
\end{equation}
where the suffix $\alpha$ describes the above processes.
We take $Y_{N_1}=0$ as an initial value of $Y_{N_1}$.
Couplings in eq.~(\ref{qlag}) and the mass of the vectorlike fermions, which are crucial for 
the determination of the PMNS matrix, are fixed to the ones used in \cite{cp}.
They are presented in the Appendix.
We assume $u=10^6$~GeV and $\rho_0=\frac{\pi}{4}$ as a VEV of the singlet scalar 
$S$ and then these couplings fix mass eigenvalues of the fourth charged lepton 
and $N_{2,3}$ as $M_E=3165$~GeV and $M_{N_{2,3}}$ given in eq.~(\ref{scotpara}), respectively.
It is noticeable that these can realize the present experimental results for 
the PMNS matrix well through a framework presented in the Appendix as described in \cite{cp}. 
We also note that the $CP$ violation, which fixes the $CP$ asymmetry $\varepsilon$, can 
take a maximal value. 

\begin{figure}[t]
\begin{center}
\includegraphics[width=7cm]{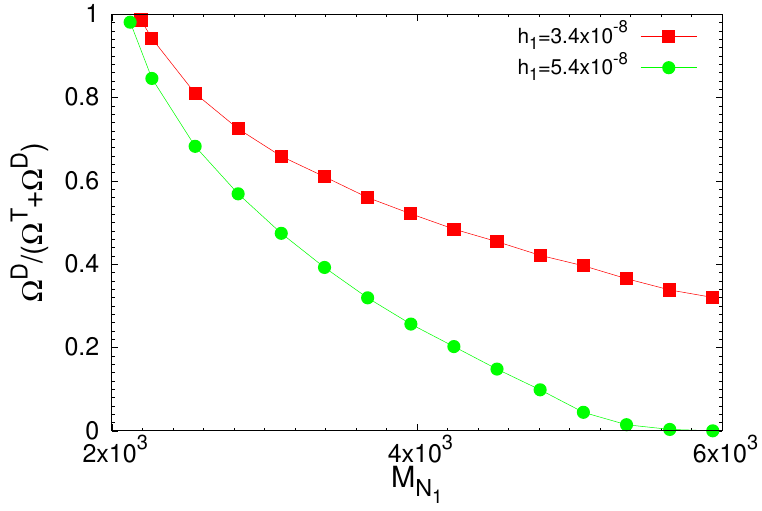}
\end{center}

\footnotesize{{\bf Fig.~4}~~The lightest right-handed neutrino mass $M_{N_1}$ dependence of the 
ratio of the relic abundance of $\eta_R^0$ originated from the $N_1$ decay 
to the total relics which are composed of the thermal one $\Omega^T$ and $\Omega^D$ from 
the $N_1$ decay. The masses of $N_{2,3}$ and $\eta_R^0$ are fixed to the ones given 
as (b) in eq.~(\ref{scotpara}). }
\end{figure}

The results of this calculation are given in Fig.~3.   
It shows that $N_1$ reaches its equilibrium number density through the introduced 
interactions and its decay produces both the sufficient lepton number asymmetry 
and a substantial part of the required relic $\eta_R^0$ abundance as in the same way as
the right panel of Fig.~1.
In this example, the $\eta_R^0$ yielded through the $N_1$ decay occupies about 
$70\%$ of the required abundance $\Omega_{\eta_R^0}h^2=0.12$.  
The remaining part is supplied from 
$\eta_R^0$ in the thermal equilibrium by setting the relevant parameters as 
$\lambda_+=0.66$ and $\lambda_3=2.6$.
Since the $N_1$ decay starts at a larger $x$ due to a smaller value of $h_1$ compared 
with the right panel of Fig.~1, its contribution to the relic $\eta_R^0$ becomes larger.

Finally, we examine how the relative share of $\eta_R^0$ produced from the $N_1$ decay 
in the total relic depends on the $N_1$ mass. 
Results are shown in Fig.~4 where the abundance of the thermal $\eta_R^0$ and the 
$\eta_R^0$ produced from the $N_1$ decay is respectively represented as $\Omega^T$ and 
$\Omega^D$, and $(\Omega^T+\Omega^D)h^2=0.12$ is imposed. 
The scalar couplings $\lambda_{3,4}$ are fixed to realize this value.
In this study, $m_{\eta_R^0}$ and $M_{N_{2,3}}$ are fixed to
the ones given as (b) in eq.(\ref{scotpara}), and then $M_{N_1}$ is allowed in the range 
2000~GeV$< M_{N_1}<$ 6000~GeV.
We also choose two values of $h_1$ which are used in the right panel of Fig.~1 and Fig.~3.  
The figure shows that the relic $\eta_R^0$ can be entirely produced by the $N_1$ decay
if $N_1$ takes a close value with the $\eta_R^0$ mass. In that case, $\lambda_3$ and $|\lambda_4|$ 
have to take large values near their perturbative bound since the $\eta_R^0$ abundance produced by 
the $N_1$ decay is large and its substantial part has to be reduced through the annihilation 
processes. We also find from the figure that $\Omega^Dh^2$ can reach the order of the required DM
abundance in a wide region of $M_{N_1}$.
This result suggests that the coincidence between the baryon number density and 
the DM density in the present Universe can be recognized as a natural consequence 
of the model. 
If $\eta_R^0$ is discovered as the DM, the origin of the relic $\eta_R^0$ can be an 
interesting subject. In that case, the present study suggests that detailed experimental 
study of the scalar couplings $\lambda_{3,4}$ might tell us the ratio of 
$\Omega^D$ to $\Omega^T$.  
 
\section{Summary}
We study a new scenario for the DM abundance in the scotogenic model
from a view point of the coincidence of the 
baryon number asymmetry and the DM abundance in the present Universe.
In this model, the abundance of the inert doublet DM is usually considered 
to be explained as the thermal relic following the WIMP scenario. 
Since it is irrelevant to the baryon number asymmetry in that case, the model 
cannot give any answer for this problem. 
However, if we note that the decay of the lightest right-handed neutrino $N_1$ 
generates both the lepton number asymmetry and the DM candidate $\eta_R^0$, 
a correlation can be found between them 
in a certain situation such that $N_1$ takes the same order mass as $\eta_R^0$.

The mass of the right-handed neutrinos can take a TeV scale value 
consistently with the neutrino oscillation data in the model. 
Moreover, the decay of such $N_1$ is known to generate the sufficient baryon 
number asymmetry through leptogenesis if the $N_1$ is in the thermal equilibrium 
through certain interactions. 
These suggest that the relic $\eta_R^0$ as the DM could be supplied not only as 
the thermal relic but also as the yields of the decay of $N_1$ which also causes 
the lepton number asymmetry. If the latter gives a dominant part of the DM 
abundance, the baryon number asymmetry has a close relation to 
the DM abundance.  
The model could give an insight for the coincidence problem.
 
We examine this idea quantitatively by solving the Boltzmann equations under 
the assumption that $N_1$ is in the thermal equilibrium initially. 
The result shows that their coincidence can be explained well. 
We also propose a well-motivated model which contains the interactions to make $N_1$ 
reach the thermal equilibrium and show how the coincidence of the baryon number asymmetry 
and the DM abundance can happen in this 
extended model. An interesting point of the model is that these interactions can give 
an explanation for the $CP$ issues in the SM. 

If both the baryon number asymmetry and a substantial part of the DM are produced 
through the decay of a common mother particle, it could be a promising scenario 
to give an answer to the coincidence problem.  
In such a context, low-scale leptogenesis may provide an interesting possibility not only from a phenomenological
viewpoint but also from a cosmological viewpoint.
The extended model studied in this paper might be considered as a prototype model 
which can realize such a scenario naturally. 
  
\section*{Appendix}
In this Appendix, we briefly address how $CP$ phases in the PMNS matrix can be induced 
through Yukawa interactions given in eq.~(\ref{qlag}) following \cite{cp}. 
If the singlet scalar gets a VEV as $\langle S\rangle=\frac{1}{\sqrt 2}ue^{i\rho_0}$, which causes
spontaneous $CP$ violation, the $CP$ phase can appear in the PMNS matrix through the 
couplings of the singlet $S$ with vectorlike charged leptons $E_{L,R}$. 
These Yukawa interactions extend the SM charged lepton mass matrix $m^e_{ij}$ to 
a $4\times 4$ mass matrix ${\cal M}_e$ such as
\begin{equation}
(\bar\ell_{L_i}, \bar E_L)\left(
\begin{array}{cc}
m^e_{ij} & {\cal G}_i \\ {\cal F}^e_j & \mu_E \\
\end{array}\right)
\left(\begin{array}{c} e_{R_j} \\ E_R \\ \end{array} \right). 
\label{lmass}
\end{equation}
where $\ell_{L_i}$ and $e_{R_i}$ are the charged leptons in the SM.
${\cal F}_j^e$, ${\cal G}_i$ and $\mu_E$ are defined as
${\cal F}_j^e=\frac{1}{\sqrt 2}(y_j^e e^{i\rho_0} + \tilde y_j^e e^{-i\rho_0})u$, 
${\cal G}_i=x_i\langle\phi\rangle$ and 
$\mu_E=\frac{1}{\sqrt 2}(y_Ee^{i\rho_0}+\tilde y_Ee^{-i\rho_0})u$.

Diagonalization of a matrix 
${\cal M}_e{\cal M}_e^\dagger$ by a $4\times 4$ unitary matrix 
$\tilde V_L$ is represented as 
\begin{equation}
\left(\begin{array}{cc} \tilde A & \tilde B \\ \tilde C& \tilde D \\\end{array}\right)
\left(\begin{array}{cc} m^em^{e\dagger}+{\cal G}{\cal G}^\dagger & 
m^e{\cal F}^{e\dagger}+\mu_E^\ast{\cal G} \\ 
  {\cal F}^em^{e\dagger}+{\cal G}^\dagger \mu_E & |\mu_E|^2 +{\cal F}^e{\cal F}^{e\dagger} \\
\end{array}\right)
\left(\begin{array}{cc} \tilde A^\dagger & \tilde C^\dagger \\ \tilde B^\dagger 
& \tilde D^\dagger \\\end{array}\right)=
\left(\begin{array}{cc}\tilde m_e^{2} & 0 \\ 0 &\tilde M_E^2 \\\end{array}\right),
\label{mass}
\end{equation}
where a $3\times 3$ matrix $\tilde m_e^2$ in the right-hand side is diagonal. 
Eq.~(\ref{mass}) requires,
\begin{eqnarray}
   && m^em^{e\dagger}+{\cal G}{\cal G}^\dagger=\tilde A^\dagger\tilde m_e^2\tilde A+
\tilde C^\dagger \tilde M_E^2\tilde C, \qquad
  {\cal F}^em^{e\dagger}+{\cal G}^\dagger\mu_E=\tilde B^\dagger \tilde m_e^2\tilde A+ 
\tilde D^\dagger \tilde M^2_E\tilde C, \nonumber \\
   && |\mu_E|^2+{\cal F}^e{\cal F}^{e\dagger}=
\tilde B^\dagger \tilde m_e^2\tilde B+ \tilde D^\dagger \tilde M_E^2\tilde D.
\end{eqnarray}  
If $|\mu_E|^2 +{\cal F}^e{\cal F}^{e\dagger}$ is much larger than each 
component of ${\cal F}^em^{e\dagger}+{\cal G}\mu_E^\ast$,
we find that $\tilde B, \tilde C$, and $\tilde D$ can be approximately expressed as
\begin{equation}
  \tilde B\simeq -\frac{\tilde A(m^e{\cal F}^{e\dagger}
+\mu_E^\ast{\cal G})}{|\mu_E|^2+{\cal F}^e{\cal F}^{e\dagger}},
 \qquad \tilde C\simeq\frac{{\cal F}^e m^{e\dagger}+{\cal G}^\dagger \mu_E}{|\mu_E|^2
+{\cal F}^e{\cal F}^{e\dagger}},
   \qquad \tilde D\simeq 1.
\end{equation}
These guarantee the approximate unitarity of the matrix $\tilde A$. 
In that case, it is also easy to find that
\begin{equation}
\tilde A^{-1}\tilde m^{e2}\tilde A= m^em^{e\dagger}+{\cal G}{\cal G}^\dagger 
-\frac{1}{|\mu_E|^2+{\cal F}^e{\cal F}^{e\dagger}}
(m^e{\cal F}^{e\dagger}+\mu_E^\ast{\cal G})({\cal F}^em^{e\dagger}+\mu_E{\cal G}^\dagger ).
\label{mix}
\end{equation}
The charged lepton effective mass matrix $\tilde m_e$ is obtained as a result 
of the mixing between the light charged leptons and the extra heavy lepton. 
If $\tilde y_j^e$ is not equal to $y_j^e$ and $|\mu_E|^2<{\cal F}^e{\cal F}^{e\dagger}$ is satisfied, 
the matrix $\tilde A$ could have a large $CP$ phase.
 
We assume that the neutrino mass matrix ${\cal M}_\nu$ is diagonalized by a tribimaximal matrix
$U_\nu$ as $U_\nu^T{\cal M}_\nu U_\nu={\cal M}_\nu^{\rm diag}$,
where the matrix $U_\nu$ can be expressed as
\begin{equation}
U_\nu=\left(\begin{array}{ccc}\frac{2}{\sqrt 6} & \frac{1}{\sqrt 3} & 0\\
\frac{-1}{\sqrt 6} & \frac{1}{\sqrt 3} & \frac{1}{\sqrt 2}\\
\frac{1}{\sqrt 6} & \frac{-1}{\sqrt 3} & \frac{1}{\sqrt 2}\\ \end{array}\right)
\left(\begin{array}{ccc}1 & 0 & 0\\
0 & e^{-i\alpha_1} & 0\\
0 & 0 &e^{-i\alpha_2} \\ \end{array}\right). 
\end{equation}
 Majorana phases $\alpha_{1}$ and $\alpha_{2}$ are written by using eq.~(\ref{theta}) as
\begin{equation}
\alpha_1=\frac{\theta_3}{2}, \quad
\alpha_2=\frac{1}{2}\tan^{-1}\left[\frac{h_1^2|\Lambda_1|\sin\theta_1+h_2^2|\Lambda_2|\sin\theta_2}
{h_1^2|\Lambda_1|\cos\theta_1+h_2^2|\Lambda_2|\cos\theta_2}\right].
\end{equation}
The PMNS matrix is obtained as $V_{PMNS}=\tilde A^\dagger U_\nu$ where $\tilde A$ is 
fixed through eq.~(\ref{mix}).
Since the matrix $\tilde A$ is expected to be almost diagonal from hierarchical masses of the 
charged leptons, the structure of $V_{PMNS}$ is considered to be mainly determined 
by $U_\nu$ in the neutrino sector.
Although tribimaximal mixing cannot realize a nonzero mixing angle $\theta_{13}$ which 
is required by the neutrino oscillation data, the matrix $\tilde A$ could compensate
this fault and a desirable $V_{PMNS}$ may be derived as $V_{PMNS}=\tilde A^\dagger U_\nu$.
If we fix the relevant parameters in eq.~(\ref{qlag}) as
\begin{eqnarray}
&&y^e=(0,3\times 10^{-3},0), \quad \tilde y^e=(0,0, 10^{-3}), \quad 
y_E=\tilde y_E= 3.3\times 10^{-6}, \nonumber \\
&&y_N=(2.2\times 10^{-3}, 6\times 10^{-3},7\times 10^{-3}), \quad 
\tilde y_N=(2.2\times 10^{-3},0,0),
\label{para}
\end{eqnarray}
$V_{PMNS}$ obtained in this way is found to be rather good realization of the 
experimental results including nonzero $\theta_{13}$ as shown in \cite{cp}. 

\newpage
\bibliographystyle{unsrt}

\begin{thebibliography}{99}
\bibitem{dmrev}G.~Jungman, M.~Kamionkowski, K..~Griest, Phys. Rep. {\bf 267} (1996) 195;
G.Bertone, D.~Hooper, J.~Silk, Phys. Rep. {\bf 405} (2005) 279.

\bibitem{pdg}R.~L.~Workman {\it et al.} (Particle Data Group), Prog. Theor. 
Exp. Phys.  2022, 083C01 (2022).

\bibitem{expb}M.~Misiaszek and N.~Rossi, Symmetry {\bf 28}, 201 (2024).

\bibitem{asymdm}K.~Petraki and R.~R.~Volkas, Int J. Mod. Phys. A {\bf 28}, 1330028 (2013).

\bibitem{scot}E.~Ma, Phys. Rev. D {\bf 73}, 077301 (2006).

\bibitem{lfv}J.~Kubo, E.~Ma and D.~Suematsu, 
Phys. Lett. B {\bf 642}, 18 (2006).

\bibitem{scotdm}R.~Barbieri, L.~J.~Hall and V.~S.~Rychkov, 
Phys. Rev. D {\bf 74}, 015007 (2006); 
M.~Cirelli, N.~Fornengo and A.~Strumia,
	Nucl. Phys. B {\bf 753}, 178 (2006); 
L.~L.~Honorez, E.~Nezri,
	J.~F.~Oliver and M.~H.~G.~Tytgat, JCAP {\bf 0702}, 028 (2007); 
Q.-H.~Cao, E.~Ma, and G.~Rajasekaran, Phys. Rev. D {\bf 76}, 095011 (2007);
S.~Andreas, M.~H.~G.~Tytgat and Q.~Swillens, JCAP {\bf 0904}, 004 (2009); 
E.~Nezri, M.~H.~G.~Tytgat and G.~Vertongen, JCAP {\bf 0904}, 014 (2009);
L.~L.~Honorez and C.~E.~Yaguna, JCAP {\bf 1101}, 002 (2011).

\bibitem{cross}T.~Hambye, F.-S.~Ling, L.~L.~Honorez and J.~Rocher, JHEP
	{\bf 0907}, 090 (2009).

\bibitem{basym}E.~Ma, Mod. Phys. Lett. {\bf A21}, 177 (2006); 
T.~Hambye, K.~Kannike, E.~Ma, and M.~Raidal, Phys. Rev. D {\bf 75}, 095003 (2007).

\bibitem{ks}S.~Kashiwase and D.~Suematsu, Phys. Rev. D {\bf 86}, 
053001 (2012); Eur. Phys. J. C {\bf 73}, 2484 (2013).

\bibitem{pmns}B.~Pontecorvo, Sov. Phys. {\bf 6}, 429 (1957); 
{\bf 7}, 172 (1958); 
Z.~Maki, M.~Nakagawa and S.~Sakata, Prog. Theor. Phys. {\bf 28}, 870 (1962).

\bibitem{spha}V.~A.~Kuzmin, V.~A.~Rubakov and M.~E.~Shaposhnikov, Phys. Lett. {\bf B155} (1985) 36.

\bibitem{fy}M.~Fukugita and T.~Yanagida, Phys. Lett. B {\bf 174}, 45 (1986).

\bibitem{otribi}P.~F.~Harrison, D.~H.~Perkins and W.~G.~Scott,
Phys. Lett. {\bf B530} (2002) 167.

\bibitem{cp}D.~Suematsu, Phys. Rev. D {\bf 108}, 095046 (2023).

\bibitem{tribi}J.~Kubo and D.~Suematsu, Phys. Lett. B {\bf 643} (2006) 336;
D.~Suematsu, T.~Toma and T.~Yoshida, Phys. Rev. D {\bf 79} (2009) 093004.

\bibitem{seesaw}P.~Minkowski, Phys. Lett. B {\bf 67}, 421 (1977);
M~Gell-Mann, P.~Ramond and R.~Slansly, in Supergravity, ed. 
by D.~Freedman and P.~Van Nieuwenhuizen, North Holland, Amsterdam, 
pp.315 (1979);
T.Yanagida, Prog. Theor. Phys. {\bf 64}, 1103 (1980);
R.~N.~Mohapatra and G.~Senjanovi\'c, Phys. Rev. Lett. {\bf 44}, 912 (1980);
J.~Schechter and J.~W.~F.~Valle, Phys. Rev. D {\bf 22}, 2227 (1980).  

\bibitem{lowlept}T.~Hugle, M.~Platscher and K.~Schmitz, Phys. Rev. D {\bf 98},
023020 (2018); D.~Borah, 
P.~S.~B.~Dev and A.~Kumar, Phys. Rev. D {\bf 99}, 055012 (2019).

\bibitem{infnonth}S.~Kashiwase and D.~Suematsu, Phys. Lett. {\bf B749} (2015) 603;
D.~Suematsu, Phys. Lett. {\bf B 760} (2016) 538.

\bibitem{infth}T.~Hashimoto and D.~Suematsu, Phys. Rev. D {\bf 102}, 115041 (2020);
T.~Hashimoto, N.~S.~Risdianto, and D.~Suematsu, Phys. Rev. D {\bf 104}, 075034 (2021);
D.~Suematsu, JCAP {\bf 08} (2023) 029.

\bibitem{rightn1}D.~Suematsu, Phys. Rev. D {\bf 100}, 055008 (2019).

\bibitem{rightn2}D.~Suematsu, Phys. Rev. D {\bf 100} 055019 (2019).

\bibitem{cp0}D.~Suematsu, Eur. Phys. J. C {\bf 81}, 311 (2021). 

\bibitem{cpasym}M.~Flanz, E.~A.~Paschos and U.~Sarkar, Phys. Lett. {\bf
	B345} (1995) 248; L.~Covi, E.~Roulet and F.~Vissani,
	Phys. Lett. {\bf B384} (1996) 169; 
A.~Pilaftsis, Phys. Rev .D {\bf 56},5431(1997);
W.~Buchm\"{u}ler and M.~Pl\"{u}macher, Phys. Lett. {\bf B431} (1998) 354.


\bibitem{dmrelic}K.~Griest, M.~Kamionkowski, and M.~S.~Turner,  Phys. Rev. {\bf 41}, 3565 (1990); 
P.~Gondolo and G.~Gelmini, Nucl. Phy. {\bf B360} (1991) 145.

\bibitem{coan}K.~Griest and D.~Seckel, Phys. Rev. D {\bf 43}, 3191 (1991).

\bibitem{xenon}E.~Aprile {\it et al.} (XENON Collaboration), Phys. Rev. Lett. 
{\bf 131}, 041003 (2023).

\bibitem{strongcp} J.~E.~Kim, Phys. Rep. {\bf 150} (1987) 1;
J.~E.~Kim and G.~Carosi, Rev. Mod. Phys. {\bf 82} (2010) 557;
D.~J.~E.~Marsh, Phys. Rep. {\bf 643} (2016) 1.

\bibitem{nb}A.~Nelson, Phys. Lett. {\bf 136B} (1984) 387; 
S.~M.~Barr, Phys. Rev. Lett. {\bf 53} (1984) 329;
A.~Nelson, Phys. Lett. {\bf 143B} (1984) 165.

\bibitem{bbp}L.~Bento, G.~C.~Branco and P.~A.~Parada, 
Phys. Lett. B {\bf 267} (1991) 95.

\bibitem{km}M.~Kobayashi and T.~Maskawa, Prog. Theor. Phys. {\bf 49}, 652 (1973).

\end{thebibliography}

\end{document}